\documentclass{aa}

\usepackage{color}

\usepackage{graphicx}
\usepackage{txfonts}
\usepackage{natbib}

\bibpunct{(}{)}{;}{a}{}{,} 
\begin{document}

\title{Interplay between heartbeat oscillations and wind outflow in microquasar IGR J17091-3624}
\author{Agnieszka Janiuk\inst{1}, 
Mikolaj Grzedzielski\inst{1},
Fiamma Capitanio\inst{2}
\and Stefano Bianchi\inst{3}
}

\institute{Center for Theoretical Physics,
Polish Academy of Sciences, Al. Lotnikow 32/46, 02-668 Warsaw, Poland \\
\email{agnes@cft.edu.pl}
\and
INAF-Instituto di Astrofisica e Planetologia Spaziali, via del Fosso del Cavaliere 100, 00133 Rome, Italy
\and
Dipartimento di Matematica e Fisica, Universit\`a degli Studi Roma Tre, via della Vasca Navale 84, 00146 Roma, Italy
}

\date{Received ...; accepted ...}
  
\abstract
{} 
{During the bright outburst in 2011, the black hole candidate IGR J17091-3624 exhibited strong quasi-periodic flare-like events (on timescales of tens of seconds)
in some characteristic states, the so-called  heartbeat state. From the theoretical point of view, these oscillations may be modeled by the process of accretion disk instability, driven by the dominant radiation pressure and enhanced heating of the plasma. Although the mean accretion rate in this source is probably below the Eddington limit, the oscillations will still have large amplitudes. As the observations show, 
the source can exhibit strong wind outflow
during the soft state. This wind may help to partially or even completely stabilize the heartbeat.}
{Using our hydrodynamical code GLADIS, we modeled the evolution of an accretion disk responsible for X-ray emission of the source. We accounted for a variable wind outflow from the disk surface. We examined the data archive from the \textit{Chandra}  and \textit{XMM-Newton} satellites to find the observed limitations on the wind physical properties, such as its velocity and ionization state. We also investigated the long-term evolution of this source, which lasted over about 600 days of observations, using the data collected by the \textit{Swift} and \textit{RXTE} satellites. During this long period, the oscillations pattern and the observable wind properties changed systematically.}
{We found that this source probably exhibits observable outbursts of appropriate timescales and amplitudes as a result of the disk instability. Our model requires a substantial wind component to explain the proper variability pattern, and even complete suppression of flares in some states. The wind mass-loss rate extracted from the data agrees quantitatively well with our scenario.}
{}

\authorrunning{A. Janiuk et al.}
\titlerunning{IGR J17091 heartbeat variability}

\maketitle

\section{Introduction}
\label{sect:intro}

The microquasar IGR J17091-3624 (hereafter IGR J17091)  is a moderately bright transient X-ray binary 
(peak flux level at $\sim$20 mCrab in the range 20-100 keV) discovered
by \textit{INTEGRAL}/IBIS in 2003 \citep{Kuulkers} and classified as a black hole candidate (BHC) for its spectral and timing properties \citep{Lutovinov03, Capitanio05,Lutovinov05}.
After the \textit{INTEGRAL} discovery, IGR J17091 was searched
in the archival data of previous X-ray missions. The data of both TTM-KVANT ~\citep{Revnivtsev}
and the BeppoSAX Wide Field Camera (WFC; ~\citet{Zand}) archives show that there were previous outbursts detected in 1994, 1996, and 2001.

The refined position of IGR J17091 provided by 
\citet{2007ATel.1140....1K}
ruled out the tentative radio counterpart previously
proposed for the source \citep{Rupen, Pandey}.
The hypothesis was supported by reanalyzing the archival radio observations that identified a faint transient radio source (sub-mJy level at 5 GHz) that showed a flux increase immediately after the 2003 outburst.
The inverted spectrum provides a signature of a compact  radio jet ~\citep{Capitanio09} that is consistent with the low/hard spectral state (LHS) observed by \textit{INTEGRAL} in the same period ~\citep{Capitanio05}.

At the end of  January 2011, the \textit{Swift}/BAT hard X-ray transient
monitor reported a renewed activity from IGR J17091. A long monitoring campaign was then carried out with \textit{Swift}/XRT, starting on February 28. Moreover, the source was extensively observed by \textit{INTEGRAL} and \textit{RXTE} . 

The 2011 outburst was by far the brightest outburst ever observed from IGR J17091, and the source flux increased up to 120 mCrab in the range 2-10 keV ~\citep{Capitanio12}.
Follow-up radio observations carried out with the ATCA telescope
measured a flat spectrum ~\citep{Corbel,Rodriguez11a, Torres} associated with self-absorbed compact jets. 
Later on, \citet{Rodriguez11a} also reported on the detection of a
discrete jet-ejection event that is typically observed when a BHC undergoes
the transition from its hard-intermediate state to the soft-intermediate state.

The \textit{RXTE}  simultaneous observation campaign revealed several  timing features, such as a 0.1-Hz quasi-periodic oscillation
(QPO), increasing in frequency
with the source flux and spectral softening ~\citep{Rodriguez,Shaposhnikov}, or
a $\sim$10 mHz QPO ~\citep{Altamirano11a}. Moreover, \textit{RXTE} /PCA data showed a continuous
progression of quasi-periodic flare-like events occurring at a rate
of between 25 and 30 mHz~\citep{Altamirano11a}. This type of variability resembles the
heartbeat variation observed in the BH binary GRS 1915+105.
~\citet{2011ApJ...742L..17A} reported a detailed study of the behavior
of the flare-like events of IGR J17091 during the first 180 days of the outburst. This study classified the different types of flares with the same scheme as used by ~\citet{Belloni00} for GRS 1915+105, which have been classified into 14 variability
classes.

In addition to all these similarities between GRS 1915+105 and
IGR J17091 ~\citep[see e. g.][for details]{Capitanio12,Pahari14}, a particularly striking difference is the X-ray flux intensity during the flare-like events.
 This fact cannot be easily explained
because unlike for GRS 1915+105, for IGR J17091 we
do not know the properties of the binary system such as the distance, the inclination angle,
BH mass, or the spin.
The GRS 1915+105 flux emission during the heartbeat states reaches the Eddington luminosity ~\citep{2004MNRAS.349..393D} and is believed to be related to disk oscillations.
As \citet{2011ATel.3230....1A}
assumed,  if the
heartbeat oscillations seen from IGRJ17091 are interpreted
as being due to the same mechanism as in GRS 1915+105, then the apparent faintness
of IGR J17091 remains unexplained unless a huge
distance or an extremely low BH mass is considered (lower than 3$M_{\odot}$).
However, the subsequent results that estimate the distance and the mass of IGR J17091 did not confirm this hypothesis.

 ~\citet{Rebusco} estimated by studying the high-frequency 
QPOs 
of the source (see also ~\citealt{Altamirano12}) a BH mass of 6M$_{\odot}$ for which the distance extrapolated from \citet{2011ATel.3230....1A} should be $>$ 22 kpc, which would mean that it lies approximateley
just outside the Galaxy. On the other hand, \citet{Rodriguez11a} calculated that the distance range is 11-17 kpc for a BH mass of 10 M$_{\odot}$, considering the transition luminosity between the hard state and the soft state ~\citep{Yu}.
Finally, the optical and near-infrared counterparts have been identified by ~\citet{Torres} during the 2011 outburst, while the results from spectral analysis indicate that the source is viewed at 
high inclination angle 
~\citep{2012ApJ...746L..20K,Capitanio12,2012ApJ...757L..12R}.
 Several hypotheses have been proposed to explain the faintness of the IGR
J17091 heartbeats, such as the spectral deformation effects due to a high inclination angle that is favored by the data ~\citep{Capitanio12}, or a low or even retrograde spin ~\citep{2012ApJ...757L..12R}.

Another peculiarity of this source is the particularly fast and ionized wind observed during the soft spectral state ~\citep{2012ApJ...746L..20K}.
  ~\citet{2012ApJ...746L..20K} reported that \textit{Chandra}  observations reveal two absorption lines at  about 6.91 and  7.32 keV. These two lines can be associated with blueshifted He-like Fe XXV and Fe XXVI, with velocities of 9000 and 15000 km/s. This projected outflow velocity is an order  of magnitude higher than previously observed in stellar-mass black holes.

In this paper, we explain the heartbeat oscillations in IGR J17091 as due to the
intrinsic oscillations of an accretion disk that is subject to a thermal-viscous instability.
As recently shown in \citet{2011MNRAS.414.2186J}, such oscillations do not require
a disk luminosity equal to the Eddington rate, and the disk instability may also occur for lower luminosities, on the order of 0.1 Eddington (the extension of the disk part that is unstable as a result
of the dominant radiation pressure decreases for low accretion rates). As we demonstrated before \citep{2002ApJ...576..908J}, the wind ejection from the
inner parts of an accretion disk may provide one possible mechanisms to partially stabilize these oscillations
and affect the variability amplitudes.
Therefore, we also investigated the relationship between the wind and the 
evolution pattern of the heartbeat oscillations.

Using our hydrodynamical code, we model the evolution of an accretion disk responsible for the X-ray emission of 
the source. We account for a variable wind outflow from the disk surface and follow the evolution of the 
oscillation phase thanks to the extensive observations campaigns performed by \textit{Swift} and \textit{RXTE}  during the 2011 
outburst of the source.
We finally compare the theoretical predictions with the constraints found with the data analysis of four observations performed by \textit{XMM-Newton} and \textit{Chandra}  satellites.

\section{Observations}
\label{sect:oservations}
\subsection{Data analysis}

IGRJ17091 was observed by \textit{Chandra} and \textit{XMM-Newton} during the 2011 outburst for a total of four pointings (two pointings for each telescope), as reported in Table~\ref{Table_to_do}.

\begin{table*}
\begin{center}
\begin{tabular}{|c|c|c|c|c|c|c|c|c|c|}
\hline
Instrument& ID &        date    & exposure & flux (2-10 keV) & source state  & wind & Radio det.  & T$_{in}$ & $\Gamma$
\\ \hline
--& -- &        --      & ks &  ergs~s$^{-1}$cm$^{-2}$ & --  & -- &-- & keV & --
\\ \hline
 XMM& 0677980201 &2011-03-27&1 &1.7$\times$10$^{-9}$& \textit{heartbeat} & no &  yes~\footnotemark[1] & 1.24$\pm$0.02 & 2.3$\pm$0.1
\\ \hline
\textit{Chandra}  &12405 &2011-08-01&31 &1.6$\times$10$^{-9}$ & \textit{heartbeat} & no  & no & 1.26$\pm$0.01 &1.9$\pm$0.1 
\\ \hline
\textit{Chandra}   &12406 &2011-11-06& 27 & 1.9$\times$10$^{-9}$  &{soft-quiet}  & yes &no & 1.7$\pm$0.1 & 2.1$\pm$0.3
\\ \hline
XMM& 070038130 &2012-09-29& 41&  0.2$\times$10$^{-9}$&hard  &no  & -- & 0.6$\pm$ 0.1 & 1.32$\pm$ 0.04~\footnotemark[2]
\\ \hline
\end{tabular} 
\caption[]{\textit{XMM-Newton} and \textit{Chandra}  observations log and spectral and timing properties. $^1$Seven days before. $^2$Cutoff at 100$^{+50}_{-30}$ keV}
\label{Table_to_do}
\end{center}
\end{table*}

The two \textit{Chandra} observations were carried out with the Advanced CCD Imaging Spectrometer (ACIS: \citealt{2003SPIE.4851...28G})
with the High-Energy Transmission Grating Spectrometer (HETGS: \citealt{2005PASP..117.1144C})
in the focal plane, and are presented in \citet{2012ApJ...746L..20K}. Data were reduced with the \textit{Chandra}  Interactive Analysis of Observations (CIAO: \citealt{2006SPIE.6270E..1VF}) 
4.6 and the \textit{Chandra}  Calibration Data Base (CALDB) 4.6.1.1 software, adopting standard procedures. First-order HEG spectra were extracted for the source and the background for a total exposure time of 31 and 27 ks for the two observations. Fits were performed on the unbinned spectra, using the \citet{1976A&A....52..307C}
statistics, to take advantage of the high spectral resolution of the gratings.
 
The \textit{XMM-Newton} observations were operated in burst and timing mode with the European Photon Imaging Camera (EPIC) CCD pn camera \citep{2001A&A...365L..18S}.
Data were reduced with SAS 13.5.0, and the screening for intervals of flaring particle background was made consistently with the choice of extraction radii, in an iterative process based on the procedure to maximize the signal-to-noise ratio described in detail by \citet{2004MNRAS.351..161P}
in their Appendix A. As a result, we extracted the source spectrum from a box region 15 (17) RAWX pixels wide, centered on the source for observation 0677980201 (070038130), while the background spectrum was extracted from RAWX columns 2--9 and 57--63. The resulting good exposure times are 1 and 46 ks. Finally, spectra were binned to oversample the instrumental resolution by at least a factor of 3 and to have no fewer than 30 counts in each background-subtracted spectral channel. The latter requirement allowed us to use $\chi^2$ statistics.

 The source was also observed up to 200 keV with the $\gamma$--ray telescopes, IBIS \citep{2003A&A...411L.131U},
onboard the \textit{INTEGRAL} satellite, simultaneously with \textit{Chandra} and \textit{XMM-Newton}.
The \textit{INTEGRAL}/IBIS  data were analyzed using the latest release of the standard Offline Scientific Analysis, OSA, version 10.0, and the latest response matrices available. The spectral analysis was focused on ISGRI, the IBIS low-energy detector \citep{2003A&A...411L.141L}.
The ISGRI spectra were extracted in the 20--200 keV energy range. A systematic error of 2\% was taken into account, as reported,
for example,  by \citet{2008int..workE.144J}.
To collect enough counts in a single spectrum, we summed contiguous IBIS spectra that showed the same shape.

In the following, errors and upper limits correspond to the 90 per cent confidence level for one interesting parameter ($\Delta\chi^2 = 2.71$), unless otherwise stated. The data were analyzed with the HEASOFT package  version v.6.15.1.

\subsection{Time variability}
\label{sect:obs1}
To determine  the timing  behavior of the source and the presence or absence of periodical flake-like events associated with the heartbeat state during a period that includes the four \textit{XMM-Newton} and \textit{Chandra}   observations (April 2011--October 2012),
 we analyzed the data  collected by \textit{Swift}/XRT  and \textit{RXTE} /PCA.
We extracted the \textit{Swift}/XRT 0.2-10 keV light curve  using the standard XRT pipeline, while for \textit{RXTE} /PCA data we used the sdt1 mode light curves. 

\begin{figure}
\includegraphics[width=\columnwidth]{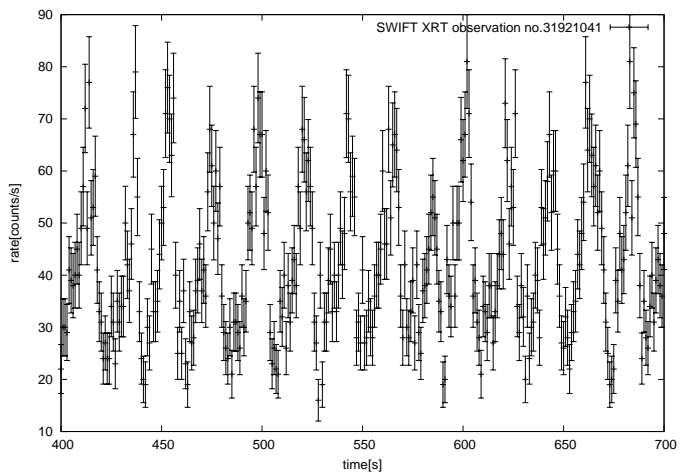}
\caption{X-ray light curve of the source IGR J17091 as observed by \textit{Swift} XRT. 
The observation ID is 31921041.
}
\label{fig:fig1}
\end{figure}

Figure \ref{fig:fig1} shows an example of  the time-variability pattern characteristic for the heartbeat state of the source. The light curve was extracted from \textit{Swift}/XRT data observed on $17^{}$ April 2011 (ID: 00031921041, exposure: 1 ks). The observation was taken in window-timing mode to avoid pile-up effects. 
Typically, IGR J17091 shows a heartbeat variability with a period of a few tens of 
seconds, as previously reported by~\citet{2011ApJ...742L..17A} and \citet{Capitanio12}.

\subsection{Wind and heartbeat detection}
\label{sect:obs2}
\begin{figure}
\includegraphics[width=\columnwidth]{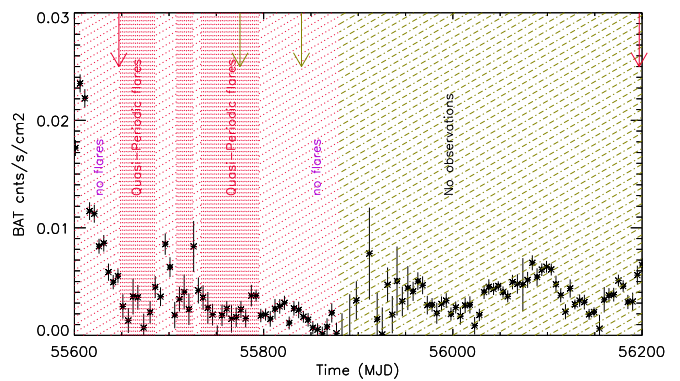}
\caption{
Swift/BAT 15-50 keV light curve superimposed on a time-line sketch created from the Swift/XRT,
RXTE/PCA, CHANDRA, and XMM data analysis, showing the anticorrelation between the wind and heartbeat in IGR J17091
 The second \textit{Chandra}  observation (second green arrow) shows a fast, ionized wind, while the first \textit{XMM-Newton} observation (first red arrows) and the first \textit{Chandra}  observation (first green arrow) lie in the heartbeat zone and do not show any detectable wind outflow. During the last \textit{XMM-Newton} observation (second red arrow), the source is in a hard state.
The parts with no flare are marked with red-oblique-dashed lines. Finally the green-oblique-dashed lines individuate periods without observations.
}
\label{fig:fiamma}
\end{figure}

Based on extensive \textit{Swift} and \textit{RXTE} observations, we report in Fig. \ref{fig:fiamma} the periods in which the periodical flare-like events have been detected, superimposed on the 15-50 keV \textit{Swift}/BAT light curve of IGR J17091. 
 ~\textit{} \textit{} 

Two simultaneous observations made by \textit{EVLA} at the time of each \textit{Chandra} pointing did not detect any radio emission \citep{2012ApJ...746L..20K}, confirming that if there is an outflow, it is probably only due to the wind ejection. However, an \textit{ATCA} observation detected a radio counterpart of IGR J17091 seven days before the first \textit{XMM-Newton} observation \citep{CorbelStefanAtel3246}. This means that a contribution of the jet emission to the outflow cannot be excluded a priori in this case.

\textit{XMM}-IBIS and \textit{Chandra}-IBIS joined spectra were modeled with the simplest possible model: an absorbed multicolor disk black body \citep{1973A&A....24..337S}
plus a power-law. Table~\ref{Table_to_do} summarize the results of the spectral fitting.

The first \textit{Chandra} observation does not show any significant absorption line in the spectrum that could be associated with a wind outflow \citep{2012ApJ...746L..20K}. The same is true for the first \textit{XMM-Newton} observation, with the exception
of a tentative detection of a feature at $\simeq7.1$ keV (see 
\citealt{2012ATel.4382....1R})
\footnote{Our analysis confirms this isolated absorption feature, detected with a nominal confidence level of $\simeq3\sigma$.}. A prominent heartbeat is present in the light curves of both observations. On the other hand, the second \textit{Chandra} observation shows a strong and fast wind, as reported by \citet{2012ApJ...746L..20K}, but no detection of any strong periodical flare-like events. Finally, the spectral analysis of the second \textit{XMM-Newton} observation indicates  that the source is in a hard state (see Table \ref{Table_to_do}): our analysis does not show any absorption line, in agreement with the general behavior of
  GBHs in this state \citep{2012MNRAS.422L..11P}.

In the next section, we study in detail the two \textit{Chandra} observations for which it is indicated that the outflow can only be ascribed to the wind ejection; the high-resolution grating spectra allow us to characterize or exclude these winds.

\subsection{Photoionization modeling of the data} 

To characterize the wind in the \textit{Chandra} observations, we prepared two tables with the photoionization code Cloudy C13.02 (last described in 
\citet{2013RMxAA..49..137F}).
We adopted the two spectral energy distributions (SEDs) described in the previous section, and, following \citet{2012ApJ...746L..20K}, a turbulence velocity $v_{turb}$=1000 km/s and an overabundance of iron by a factor of two (the other chemical abundances are as in Table 7.1 of the Cloudy documentation).

\begin{figure}
\includegraphics[width=\columnwidth]{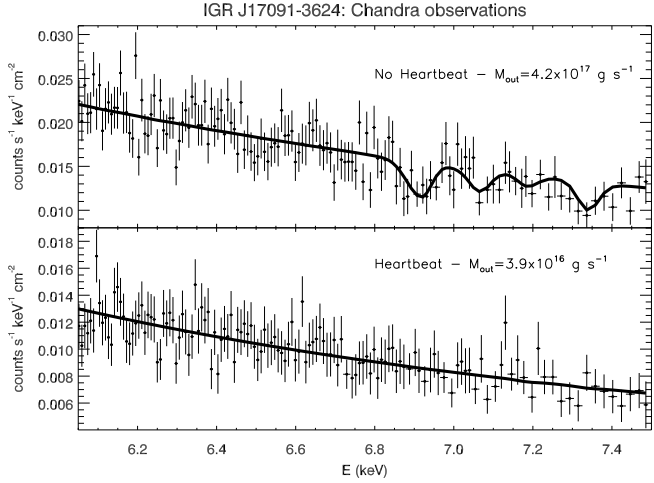}
\caption{\textit{Chandra} ACIS-S HETG spectra for observations 12406 (\textit{upper panel}) and 12405 (\textit{lower panel)} in the 6-7.5 keV energy range. The wind components observed in observation 12406 are completely undetectable in observation 12405 if we assume that the wind density is a factor of 10 lower. See text for details.}
\label{winds}
\end{figure}

We first fitted the second \textit{Chandra} observation with a continuum model resembling that presented in \citet{2012ApJ...746L..20K}, that is, a disk black-body and a power law. Two photoionized components are required to fit the absorption residuals around 6-7 keV: one with an outflow velocity $v_1=9\,700^{+800}_{-700}$ km s$^{-1}$, $\log\xi_1=3.4^{+0.2}_{-0.3}$, and $\log\mathrm{N_{H1}}=22.1^{+0.2}_{-0.4}$; the other with an outflow velocity $v_2=15\,700\pm600$ km s$^{-1}$, $\log\xi_2=3.8^{+0.2}_{-0.1}$, and $\log\mathrm{N_{H2}}=22.5\pm0.3$. We note here that these values perfectly agree with those reported by \citet{2012ApJ...746L..20K}. The spectrum and best-fit model are shown in Fig.~\ref{winds}.

From the best-fit values of the ionization parameter $\mathrm{\xi=L_{ion}/nr^2}$, assuming the ionizing (1-1000 Ryd) luminosity as in the input SED of $\mathrm{L_{ion}}=3.7\times10^{37}$ erg s$^{-1}$, we derive $n_1r_1^2\simeq1.5\times10^{34}$ and $n_2r_2^2\simeq5.8\times10^{33}$ cm$^{-1}$ for the two wind components. The corresponding (spherical) mass outflow rates would be
\begin{equation}
\dot{M}_{wind} \simeq 1.23 m_p f v \Omega nr^2
,\end{equation}
where $m_p$ is the proton mass (the correction factor of 1.23 comes from assuming cosmic elemental abundances), $\Omega$ is the covering factor, and $f$ the filling factor. The absence of emission lines suggests $\Omega/4\pi\simeq0.5$ \citep{2012ApJ...746L..20K}, so our estimated mass outflow rates are
\begin{align}
\dot{M}_{wind1} \simeq 1.8\times10^{20} \,f_1\,\,\mathrm{g\, s^{-1}}\\
\dot{M}_{wind2} \simeq 1.2\times10^{20} \,f_2\,\,\mathrm{g\, s^{-1.}}
\end{align}
The filling factors f${_1}$ and f${_2}$ of the wind components are more difficult to estimate. It is reasonable to assume that the extension of the wind is similar to its launching radius, $r\simeq\Delta r$, so that the observed absorption column densities are related to the filling factors, which means that $N_H=fnr$.  To break the degeneracy between $n$ and $r$, we assumed that the outflow velocity of the wind is on the order of the escape velocity at its launching radius. This leads to $r_1\simeq1900$ r$_g$ and $r_2\simeq760$ r$_g$, corresponding to $n_1\simeq5.1\times10^{15}$ cm$^{-3}$ and $n_2\simeq1.3\times10^{16}$ cm$^{-3}$. The resulting filling factors, $f_1\simeq0.0015$ and $f_2\simeq0.0037$, give $\dot{M}_{wind1}\simeq2.7\times10^{17}$ g s $^{-1}$ and $\dot{M}_{wind2}\simeq4.2\times10^{17}$ g s $^{-1}$.

The first \textit{Chandra} observation does not show any absorption feature. If we assume that the density of the wind is lower by a factor of 10 with respect to the density characterizing the observed wind in the second observation, we would expect $\xi$ and $N_H$ to re-scale linearly (but we also take into account the small difference in the SED, and, therefore, in L$_{ion}$). As shown in Fig.\ref{winds}, the two wind components would be undetectable in our data if they had these
parameters. Assuming that all the other properties of the wind remain the same, we obtain lower mass outflow rates of $\dot{M}_{wind1}\simeq2.5\times10^{16}$ g s $^{-1}$ and $\dot{M}_{wind2}\simeq3.9\times10^{16}$ g s $^{-1}$.

\section{Model of the radiation pressure instability}
\label{sect:sect2}

The outbursts of accretion disks may be induced by the two main types of 
instabilities that lead to the thermal-viscous oscillations. These are 
the radiation pressure instability and the partial hydrogen ionization
instability, both known for over 40 years in theoretical astrophysics. 
For a detailed discussion of these instabilities we refer to \citet{
2011MNRAS.414.2186J}
and to the literature quoted therein.
We here studied the
accretion disk instability induced by the dominant radiation pressure 
and modeled it to explain the behavior of the microquasar IGR J17091.

The stationary, thin accretion disk model in classical theory is based on an
$\alpha$ prescription for the viscous energy dissipation.
In the $\alpha$-model we assume the nonzero component $T_{r \phi}$
of the stress tensor to be proportional to the total pressure.
The latter includes the radiation pressure component, which scales 
with temperature as $T^{4}$ and blows up in hot disks for high accretion rates.
This in turn affects the heating and cooling balance between the energy 
dissipation and radiative losses.

This balance, assuming hydrostatic equilibrium, is
calculated numerically with a closing equation for the
locally dissipated flux of energy given the black hole mass and 
global accretion rate. The local solutions may be conveniently plotted 
on the so-called stability 
curve. This {\it S-shaped} scheme represents an 
annulus in the accretion disk,
 with temperature and surface density determined by the
accretion rate. If the accretion rate is low, the 
annulus of the disk is dominated by the gas pressure and stable. For higher 
accretion rates, the annulus
is dominated by radiation pressure and unstable. An upper stable branch of this curve 
appears for even higher accretion rates, close to the Eddington limit,
and is due to efficient advective cooling.
The higher the global accretion rate, the more annuli of the disk will be
affected by this instability. 
In consequence, for a larger size of the instability zone, the disk will undergo
stronger amplitude oscillations. Below a critical accretion rate, the instability zone completely vanishes 
and the whole disk is dominated by gas pressure and hence stable.
We determined this rate to be about  3.6 per cent Eddington.

\subsection{Physics of the model}
\label{sect:physics}

\subsubsection{Equations}

We assumed nondiagonal terms in stress tensor in cylindrical coordinates proportional to the total pressure with a constant viscosity $\alpha$, as introduced by 
\citet{1973A&A....24..337S},
 
\begin{equation}
T_{r \phi} = - \alpha P,
\label{pirphi}
\end{equation}
where $P$ is the pressure.

If we consider a thin disk whose energy loss is only connected with horizontal flows, local energy loss in the disk per unit volume per unit time
 is given by
\begin{equation}
\frac{\partial F}{\partial z} = T_{r \phi} {d \Omega \over d r}
\label{flux}
,\end{equation}
where $v_\phi$ is the angular (Keplerian) velocity in the disk
\begin{equation}
\Omega_K = \sqrt{\frac{GM}{r^3}}
.\end{equation}
It is convenient to express the quantities in terms of Eddington units, 
and the dimensionless accretion rate is
\begin{equation}
\dot{m}=\frac{\eta \dot{M}c^{2}}{ L_{Edd}}=\frac{\dot{M}}{\dot M_{Edd}}
.\end{equation}
Here $\eta$ is the efficiency of accretion, equal to $1/16$ in the pseudo-Newtonian approximation.
According to Eqs. \ref{pirphi} and \ref{flux}, the final formula for the viscous energy dissipation is
\begin{equation}
\frac{\partial F_{tot}}{\partial z} = \alpha P {d \Omega \over d r}.
\end{equation}
The equilibrium pressure consists of gas and radiation pressure,
\begin{equation}
P = P_{\rm gas} + P_{\rm rad}
,\end{equation}
where we assume that gas is an ideal gas of protons of mass $m_{p,}$
\begin{equation}
P_{\rm gas} = \frac{\rho k_B}{m_p} T
,\end{equation}
and the radiation pressure is given by
\begin{equation}
P_{\rm rad} = \frac{4 \sigma_B}{3 c} T^{4}
.\end{equation}

We consider here the model of a vertically averaged disk (effectively,
a 1.5-dimensional model, as the motion in the angular direction is also accounted for).
In the simplest possible stationary case, we close the equations by an energy conservation equation
\begin{equation}
F_{\rm tot}=Q_{+} = Q_{-}
\label{encor}
,\end{equation}
where the total flux locally 
emitted from the surface of the accretion disk is given by the global
parameters\begin{equation}
F_{\rm tot} = {3 G M \dot M \over 8 \pi r^{3}} f(r)
\label{eq:ftot}
,\end{equation}
where $f(r)$ is given by the boundary condition on the inner edge of accretion disk.
The vertically averaged viscous heating rate in the disk is
\begin{equation}
Q_{+} = \frac{3}{2} \alpha P H \Omega_{K}
,\end{equation}
and the radiative cooling rate per unit time per surface unit is
\begin{equation}
Q_{-} = \frac{4 \sigma_{B} T^{4}}{3 \kappa \Sigma},\end{equation}
where $\kappa$ is the electron scattering opacity,
which is equal to 0.34  cm$^{-2}$ g$^{-1}$.

The advective cooling is included in the stationary model as
\begin{equation}
Q_{-} = F_{tot}(1-f_{adv}),
\end{equation} 
where $f_{adv}$ is defined as follows \citep{1982AcA....32....1M,1988ApJ...332..646A,2002ApJ...576..908J}:
\begin{equation}
f_{adv} = \frac{ - 2 r P}{ 3 \rho GM f(r)}  q_{adv}
,\end{equation}
and $q_{adv}$ is given by
\begin{equation}
q_{adv} = (12 - 10.5 \beta ) \frac{d \ln T}{d \ln r} - (4-3 \beta) \frac{d \ln \rho}{d \ln r}
,\end{equation} 
where $\beta=P_{\rm gas}/P$.
The fraction $q_{adv}$  in stationary model is assumed to
bea constant on the order of unity.

\subsubsection{Time-dependent 1.5-D hydrodynamics}

In the time-dependent model we solve the full set of equations of hydrodynamics. 
We consider a thin-disk model, so in the conservation of mass (continuity) equation we neglect the z-behavior of density and velocity fields: 
\begin{equation}
\frac{\partial ( \Sigma v_{r})}{\partial r} + r \frac{\partial \Sigma}{\partial t} =0
\label{con}
.\end{equation}
The mass conservation equation is therefore
\begin{equation}
\frac{\partial \Sigma }{\partial t} = \frac{1}{2 \pi r}  \frac{\partial \dot{M}}{\partial r}
,\end{equation}
with
\begin{equation}
\dot{M} = - 2\pi r \Sigma v_r
.\end{equation}
The angular momentum conservation is included in the equation
\begin{equation}
\dot{M} \frac{d}{dr} (r^2 \Omega) = - \frac{\partial}{\partial r} (2 \pi r^2 T_{r \phi})
\label{eq:dotm}
.\end{equation}
We define $\nu$ as the kinematic viscosity
\begin{equation}
 T_{r \phi} = \alpha P H = { 3 \over 2} \Omega \nu \Sigma
 \label{eq:nudef}
.\end{equation}
From the mass and angular  momentum conservation Eq. \ref{eq:dotm} 
we obtain the final formula on the evolution of the surface density of disk:
\begin{equation}
\frac{\partial \Sigma}{\partial t} = \frac{1}{r} \frac{\partial }{\partial r}( 3 r^{1/2} \frac{\partial }{\partial r}( r^{1/2} \nu \Sigma ))
.\end{equation}
It has the form of a simple diffusion equation when we transform the variables
\begin{equation}
\frac{\partial \Xi }{\partial t}  = \frac{12}{y^2} \frac{\partial^2 \Xi}{\partial y^2} (\Xi \nu)
,\end{equation}
where $r = 2 y^{1/2}$ and $\Xi = 2 r^{1/2} \Sigma$.
The radial velocity is given by
\begin{equation}
v_r = - \frac{3}{\Sigma} r^{-1/2} \frac{\partial}{\partial r}(\nu \Sigma r^{1/2})
.\end{equation}
To close the set of the equations, we need the energy conservation equation.
In the disk we may have radiative cooling as well as convection, advection, and diffusion.
In our model we neglect the effects of convection and heat diffusion and assume the same equation as in \citet{2002ApJ...576..908J},
\tiny{
\begin{equation}
\frac{\partial \ln T}{\partial t} + v_r \frac{\partial \ln T}{\partial r} = \frac{4 - 3 \beta }{12 - 10.5 \beta} (\frac{\partial \ln \Sigma}{\partial t} - \frac{\partial \ln H}{\partial t} + v_r \frac{\partial \ln \Sigma}{\partial r}  ) + \frac{Q_{+} - Q_{-}}{(12-10.5 \beta)PH}
.\end{equation}

}
\normalsize
\subsection{Numerics}
\label{sect:numerics}

We used the code global accretion disk instability (GLADIS),
whose basic framework was initially described by 
\citet{2002ApJ...576..908J}. 
The code was subsequently developed 
and applied in a number of works to model the evolution of accretion disks
in Galactic X-ray binaries and AGN
\citep{2005MNRAS.356..205J, 2009ApJ...698..840C}.
The current version includes the parallelization through the message-passing interface 
method (MPI) \citep{2012A&A...540A.114J}
and allows for computations with a variable time-step down to 
a thermal timescale, adjusting to the speed of local changes of the disk structure.

\subsubsection{Grid and regularity}

First, we tested the code parameterized with the accretion rate equal to the Eddington rate, that is, $\dot m = 1$, and a central object mass of $M = 10 M_{\odot}$. The grid covers the area from the inner radius $R_{in}$ equal to $3$ Schwarzschild radii to external radius equal to $100$ Schwarzschild radii.

\begin{table*}
\begin{center}
\begin{tabular}{|c|c|c|c|c|c|c|c|}

\hline
$N_{p}$ &       period  & error &  variance & duration of outburst & error & variance & $N_{o}$
\\ \hline
25 &    441.16 &        52.7 &  341.63 &        10.68 & 0.38  & 2.52 &  44

\\ \hline 30 &  684     & 77.9  & 397.3  &      29.28   &3.81 & 19.81 & 27
\\ \hline 40 &  1199    & 47.6  & 184.4 &       57.06   & 2.81 &        11.25   & 16
\\ \hline 50 &  1342    & 43.8  & 158.0 &       70.66   & 3.45 &        12.89   & 15
\\ \hline 75 &  1381    & 21.1  & 76.0 &        76.67   & 2.11 &        7.89    & 14
\\ \hline 100 & 1402    & 18.5  & 64.2 &        78.57   & 2.06 &        7.42    & 13
\\ \hline
\end{tabular} 
\caption[]{Resulting heartbeat outbursts modeled with different grids. The first column, $N_{p}$, gives the number of grid points. The last column, $N_{o}$, gives the number of outbursts in the time interval of 20 ks. The other quantities are given in seconds.}
\label{t1}
\end{center}
\end{table*}

In Table \ref{t1} we  summarize the tests of our numerical grid.
We computed the average value $\bar{T}$ in our modeled light
curve as
\begin{equation}
 \bar{T} = \frac{1}{N} \sum\limits_{i=1}^N T_i
,\end{equation}
  where $T_{i} = t_{i+1}-t_{i}$ are distances between $i+1^{th}$ and $i^{th}$ outburst peaks. The variance $V_T$ of periods for each grid is
\begin{equation}
 V_T = \sqrt{\frac{1}{N} \sum\limits_{i=1}^N (T_i - \bar{T})^{2}}
,\end{equation}
and errors are given as
\begin{equation}
 E_T = \sqrt{\frac{1}{N(N-1)} \sum\limits_{i=1}^N (T_i - \bar{T})^{2}}.
\end{equation}

We note that a denser grid leads to a better regularity of outbursts. For a number of grid points equal to $100$ the average period is equal to 1401.92 s, with variance of periods between outbursts 
equal to 64.1 s, which is less that 5 per cent. We conclude that outbursts are regular in our model and accurately describe the heartbeat effect.

\subsection{Parameters}
\label{sect:parameters}

We adopted the global parameters for the accretion disk model to be the
black hole mass of IGR J17091 equal to 6 $M_{\odot}$, as estimated by \citet{Rebusco} 
and (arbitrary) viscosity parameter $\alpha=0.1$.
To determine the accretion rate, we determined the distance of IGR17091 using the same method as reported by ~\citet{Rodriguez11a} rescaled for 6 $M_{\odot}$, 
which gives a distance range of $d=8-13$ kpc. The absolute luminosity range based on this 
distance range is $L=3.1 \times 10^{37} - 8.2 \times 10^{37}$ erg s$^{-1}$. 
For the black hole mass 
adopted here this gives a dimensionless accretion rate range of $\dot m=0.041 -0.108$,
which is equal to $\dot M = 0.86 - 2.12 \times 10^{-8} M_{\odot}$ yr$^{-1}$ 
for this source.

\subsection{Results}
\label{sect:results}

The radiation pressure instability model with the parameters given above leads
to the pronounced, regular luminosity outbursts.
Obviously, the heating proportional to the gas pressure
results in a completely stable accretion disk, with a constant luminosity, 
regardless of the accretion rate.
Moreover, 
if we assume the viscous heating
given by the geometrical mean of the gas and radiation pressures, no outbursts
occur.
 Therefore, to model the heartbeat states of IGR J17091, we 
need the total pressure prescription for the viscous torque (Eq. 23).

\subsubsection{Behavior of the disk during an outburst cycle}
\label{sect:behaviour}

If there were no stabilizing mechanism, the radiation-pressure dominated disk would not survive. 
This is because in these parts of the disk the 
decreasing density leads to the temperature growth. 
In consequence, the local accretion rate increases and more material is 
transported inward. The disk annulus empties because of both the increasing 
accretion rate and decreasing density, so there is no self-regulation 
of the disk structure.
However, the so-called slim-disk solution, where advection of energy
provides an additional source of cooling in the highest accretion rate regime
(close to the Eddington limit), acts as a stabilizing branch. 
Therefore, 
even if a large part of the disk is dominated by radiation pressure, 
advection of some part of energy allows the disk to survive and oscillate 
between the hot and cold states. This oscillating behavior leads to periodic 
changes of the disk luminosity, as shown, for instance, in \citet{2000ApJ...542L..33J}
for the 
black hole X-ray binary disks.

\begin{figure}
\includegraphics[width=7cm]{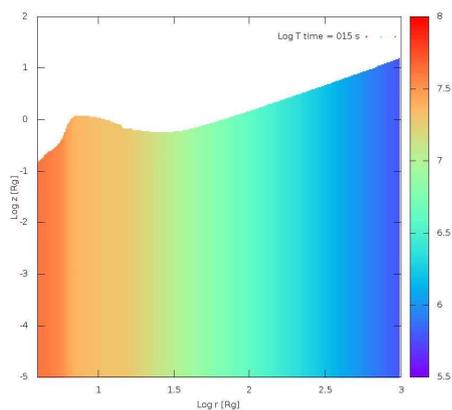}
\includegraphics[width=7cm]{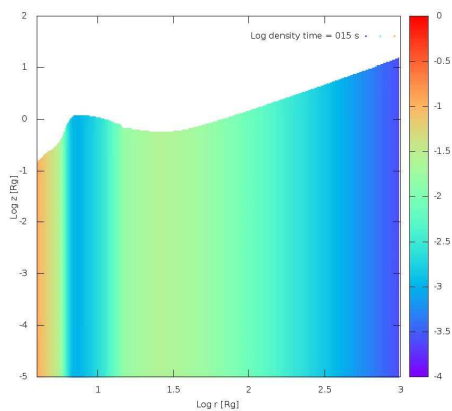}
\caption{Profiles of the disk density and temperature during
an outburst in cgs units. 
The distribution is vertically averaged (note that the approximation of a geometrically thin disk also holds during the flare). 
Parameters: viscosity $\alpha=0.1$, 
 mean accretion rate $\dot{m}$=0.1, and black hole mass 
$M=6 M_\odot$. Distance is expressed in Schwarzschild radius units.}
\label{fig:profiles}
\end{figure}

In Fig. \ref{fig:profiles} we show the distributions of density and temperature
in the accretion disk in the peak of an outburst. The quantities are vertically averaged, 
while we also computed the disk thickness, as indicated on the y-axis. The ratio of $H/r$ is always lower than about 0.1, which justifies our approach.
In the outburst peak, the inner disk becomes flaring and its temperature rises, 
while the density drops by an order of magnitude.

\begin{figure}
\includegraphics[width=\columnwidth]{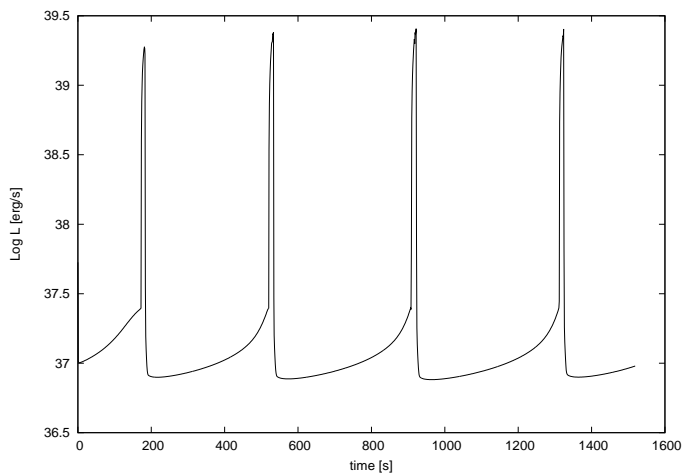}
\caption{Example of light curve modeled with an accretion disk instability, 
for $A$=0 (without wind), a mean accretion rate $\dot{m}$=0.1, and a black hole mass 
$M=6 M_\odot.$}
\label{fig:lightcurve1}
\end{figure}

Figure \ref{fig:lightcurve1} shows an exemplary light-curve, 
with oscillations on a timescale of a few hundreds of seconds, as calculated for the 
black hole mass and accretion rate given in Sec. \ref{sect:parameters}.
The amplitudes of these outbursts, however, are much larger than those 
observed in the source's X-ray light-curves, and by over an order of magnitude 
exceed the luminosity changes detected in IGR J17091.

\subsubsection{Partial stabilization of the disk outbursts as
a result of wind outflow}
\label{sec:wind_general}

As known before (e.g., \citealt{2005MNRAS.356..205J}, see also \citealt{2012MNRAS.421..502N}), 
the disk oscillations may be partially stabilized by
the wind outflow and/or corona. Here we suggest that
part of the energy generated in the accretion process is used to accelerate the
particles to eject the outflow from the disk surface. The reduced heating will 
result in smaller amplitude outbursts produced through the instability.

Following \citet{2002ApJ...576..908J},
we corrected for the fraction of energy that is transferred to the disk surface 
(so in principle, we modified the conditions of the vertical energy transport 
that are behind the vertically integrated model). We adopted a 
function that assumes wind-launching power scales
quadratically with the local accretion rate:
\begin{equation}
f_{out} = 1 - {{1} \over {(1+A {\dot m}^{2})}}.
\label{eq:fout}
\end{equation}
Various other mathematical prescriptions were studied, for example,
in \citealt{2000ApJ...535..798N}, who found no qualitative importance
on the details of this function.
This scaling represents the physical nature of the process and 
simply means that the wind particles are ejected when
the local disk luminosity instantaneously 
approaches the Eddington limit: $\dot m = \dot M / \dot M_{Edd}  = L/L_{Edd}$.
Here $f_{out}$ is a dimensionless number in the range (0,1), and in our computations it 
depended on both radius and time via $\dot M \propto r \Sigma v_{r}$.
This fraction modifies the local heating-to-cooling balance, which means that effectively  $f_{out}$ enters into the denominator in 
Eq. \ref{eq:qminus}.
The free constant $A \ge 0 $ is our model parameter; it 
describes the strength of the wind and its mass-loss rate. 
Note that in the evolutionary calculations, 
we did not update the local density of the disk according to this mass loss because the latter 
is lower by many orders of magnitude.

We first phenomenologically determined the value of the wind strength parameter $A$ ,
so that the resulting reduced outburst amplitudes with nonzero wind outflow
reproduce the observed variability scale of the microquasar IGR J17091, like those
 shown in
Fig. \ref{fig:fig1}.
\begin{figure}
 \begin{center}
\includegraphics[width=\columnwidth]{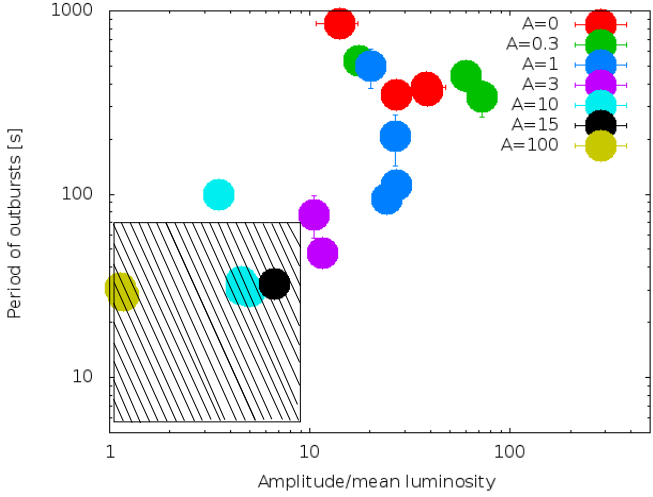}
\end{center}
\caption{Dependence of the outburst amplitudes and periods on the wind strength $A$. Shaded regions mark observed amplitudes and timescales of the microquasar heartbeats. 
Different colors mark different values of the parameter, as indicated in the plot. Models were ran for $\dot{m}$ between $0.04$ and $0.1$.
In general, the lower the accretion rates, the shorter the outbursts, 
as long as the instability occurs.}
\label{fig:wind_params}
\end{figure}

Figure \ref{fig:wind_params} shows the dependence of outburst amplitudes on the
changing wind strength parameter. Different symbols in this figure mark different values of $A$, and we show several results for each $A$, 
depending on the adopted mean accretion rate, $\dot m$, 
as taken in the range of
values determined for microquasar IGR J17091. Values lower than $A=10$ were 
excluded because they yielded too large amplitudes and long outburst
timescales 
for any plausible  accretion rate. On the other hand, values of 
$A>25$ yield too small amplitudes, and outbursts cease completely for $A=100$.

\begin{figure}
\begin{center}
\includegraphics[width=\columnwidth]{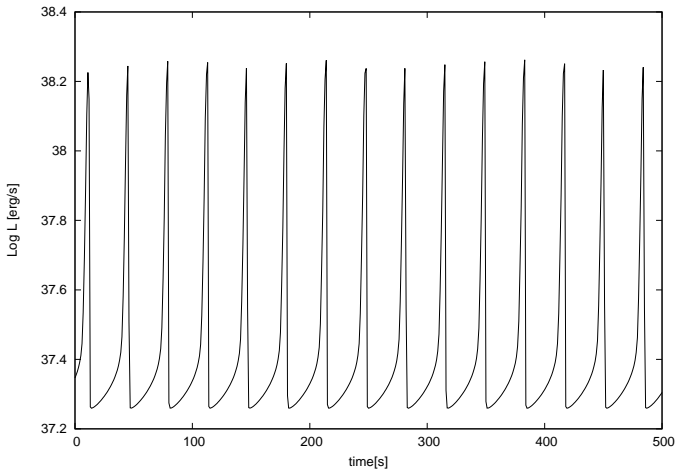}
\end{center}
\caption{Model light-curve of the accretion disk. The luminosity 
changes are due to the radiation pressure instability.
The wind strength is $A=15$. Other parameters:
black hole mass $M= 6 M_{\odot}$, viscosity $\alpha=0.1$, accretion
rate $\dot M = 2.12\times 10^{-8}$ $M_{\odot}$yr$^{-1}$ ($\dot{m} = 0.1$).}
\label{fig:lightcurve2}
\end{figure}

Figure \ref{fig:lightcurve2} shows an exemplary light curve from the model with wind,
which best represents the behavior of our source in its heartbeat state.
Since the accretion disk emission flux peaks at about 10 Schwarzschild radii, 
the light curve shown here corresponds to the soft X-ray band for a stellar-mass 
accreting black hole.

\subsubsection{Average mass loss, density, and column density of the wind }
\label{sec:massloss}

The mass-loss rate in the vertical direction is equal to the ratio of the locally generated flux
in the accretion disk, with a fraction determined by $f_{out}$,
 to the energy change per particle, $\dot m_{z} = F_{tot} f_{out}/(\Delta E/m_{p})$. 
The local flux is given by Eq. (\ref{eq:ftot}) of the standard accretion disk theory,
and we assumed that this relation also holds in the hydrodynamical computations.
The energy change per particle is on the order of the virial energy, $\Delta E =  B k T_{vir} = B GM m_{p}/r$, with $B\sim 1$,
so that expressing  the mass-loss rate in terms of local variables, in the units of 
[g s$^{-1}$ cm$^{-2}$], we have 
the final formula
\begin{equation}
\dot{m}_z(r) = B^{-1} \frac{3}{4} \frac{1}{r} \Sigma v_r f(r) f_{out}
\label{dotmz}
,\end{equation}
with $f_{out}$ given by Eq. (\ref{eq:fout}).

To obtain the total mass loss in the wind, we integrated $\dot m_z$ over the disk:
\begin{equation}
\dot{M}_{wind} = \int_{R_{min}}^{R_{max}} \dot{m}_z 4 \pi r dr
.\end{equation}
We assumed the outer radius of the accretion disk to be equal to $4\times 10^{4} R_{g}$, which is
the largest possible size of the disk in IGR J17091, as estimated from
the supposed parameters of the binary.
The smallest radius, $R_{\rm min}$, could be located at the marginally stable orbit
around the black hole, that is, at 3 $R_{g}$ for a nonrotating hole, or anywhere 
above it. We therefore list several results here for the smallest radii from which the unbound 
wind is launched. 

The actual mass loss will be on this order, or somewhat smaller, as the wind particles may
be accelerated to obtain kinetic energy not necessarily equal to the virial energy, but rather
to have a velocity a few (2-3) times their escape velocity at the radius r. Here we therefore calculate the upper limit for the wind column density, depending on 
our model parameter $A$.

The wind mass-loss changes in time as a result of the outbursts. Averaging the solutions, 
we obtain the total mass loss caused by the wind at the level of 
$\dot{M}_{\rm wind} =  3 \times 10^{16} - 2 \times 10^{17}$ g s$^{-1}$. 

We calculate the wind density at $R_{\rm max}$
\begin{equation}
\rho_{\rm wind} = \frac{\dot{M}_{wind}}{4 \pi R_{max}^2 v_{esc}}
\label{rhowind}
,\end{equation}
where $v_{\rm esc}$ is the escape velocity at $ R_{\rm min}$. 
In our best-fit heartbeat model, the density of the wind is equal to 
$\rho_{\rm wind} = 2.0\times 10^{-16} - 7.2\times 10^{-16}$ g cm$^{-3}$, depending on the wind-launching radius.

Assuming that wind consists of protons, we calculate the column density of the (spherically symmetric) wind:
\begin{equation}
\int \rho_{\rm wind}(\vec{r}) dr = \rho_{\rm wind}(R_{\rm max}) R_{\rm max}.
\end{equation}
The observable column number density of particles is then be given by
$N_{\rm H} = {\rho_{\rm wind}} R_{\rm max}(f m_{\rm p})^{-1}$
, where $f$ is the filling factor, in principle unknown and constrained from the wind observations.
The resulting wind mass-loss rate and density for an assumed wind strength 
parameter $A=15$ and several values of 
the inner radius at which the wind is launched
are summarized in Table \ref{tajet15}. 

\begin{table}
\begin{center}
\begin{tabular}{|c|c|c|c|}
\hline
$R_{min}$ & $\dot{M}_{\rm wind}$ [g s$^{-1}$] &  $\rho_{\rm wind}$ [g cm$^{-3}$] & $N_{\rm H}$ [cm$^{-2}$] 
\\ \hline $3$ &         $2.18 \times 10^{17}$ &   $2.0 \times 10^{-16}$   &       $5.66 \times 10^{21}$  
\\ \hline $70$ &        $8.84 \times 10^{16}$ &   $3.9 \times 10^{-16}$   &      $1.10 \times 10^{22}$ 
\\ \hline $150$ &       $7.68 \times 10^{16}$ &   $5.0 \times 10^{-16}$   &      $1.07 \times 10^{22}$ 
\\ \hline $350$ &       $6.04 \times 10^{16}$ &   $7.1 \times 10^{-16}$   &      $2.00 \times 10^{22}$ 
\\ \hline $2000$ &      $3.06 \times 10^{16}$ &         $7.2 \times 10^{-16}$   &     $2.04 \times 10^{22}$ 
\\ \hline
\end{tabular} 
\caption{Exemplary results of the disk/wind model. The parameters are 
the accretion rate $\dot m = 0.1$, wind strength $A=15$, black hole mass $M = 6 M_{\odot}$, and wind extension 
$R_{\rm max} = 40000 R_{g}$. The column density is calculated assuming $f_{1}=0.0015$.}
\label{tajet15}
\end{center}
\end{table}

\subsection{Comparison between the wind model and observations of IGR J17091}

To reconcile the model with observations of the wind in IGR J17091, we need to constrain the wind launching zone in its radial extension
to specify $R_{min}$ and $R_{max}$.
We also need to determine
the wind strength parameter $A$ to either reproduce the amplitude of the observed heartbeat oscillations
or supress them completely. 
For radii below $\sim 70$ Schwarzschild radii, the 
dependence of $\dot m_{z}(r)$ derived form Eq. \ref{dotmz} is a complicated function, while at larger radii it scales
with radius as a simple power law with index $-1.8$. Therefore, we fitted this relation and its normalization to our 
best-fit model for the heartbeat oscillating disk, that is, with $A=15$. After simple integration, we obtain that
\begin{equation}
\dot M_{\rm wind}(r, R_{\rm max}) = 1.635\times 10^{16} (R_{\rm max}^{0.2}-r^{0.2}) ~~ {\rm [g ~ s^{-1}]}
,\end{equation}
where in the above the argument $r=R_{min} \ge 70$ and is given in the units of $2GM/c^{2}$, and we already assumed the covering factor of $\Omega=2\pi$.

From this relation, we can constrain the extension of the wind-launching zone, taking into account
the radius $r$ determined by the velocity of the wind equal to the escape velocity at that radius,
which can be constrained by the spectral analysis (provided the wind is detectable in the data).
Therefore, if we adopted the outflow velocities to be the same as in the 
second \textit{Chandra} observation, then we can tentatively estimate that
for the first wind component with a mass-loss rate of about $\dot M_{\rm wind1}=2.5 \times 10^{16}$ g s$^{-1}$, 
the extension of the launching zone should be up to $R_{\rm max}\approx 4900$ Schwarzschild radii.
The second wind component, with $\dot M_{\rm wind2}= 3.9 \times 10^{16}$ g s$^{-1}$, should be launched 
in the zone up to $R_{\rm max}\approx 5900$.

Furthermore, we note the strong dependence of the wind mass-loss rate on the strength 
parameter $A$.

With $A=300$, a stable disk solution is found in our simulations, 
while the mass-loss rate
$\dot M_{\rm wind}$ is about $\sim 12$ times higher than for
our
heartbeat model with $A=15$.
Such a strong wind 
definitely stabilizes the heartbeat oscillations, while 
it should give clear observable signatures in the \textit{Chandra} spectra, such as those presented in Fig. \ref{winds}.
Taking the values of the wind velocities and mass-loss rates to be the same as those in the first \textit{Chandra} observation, we obtain observable wind-laucching zones
of $950-4200$ and $380-4700$ Schwarzschild radii for the first and second wind components.

\section{Discussion}
\label{sect:discconc}

\subsection{General picture for IGR J17091}

The photoionization modeling of winds detected during the two \textit{Chandra} observations revealed a strong wind in the state without 
heartbeat oscillations. We found two compoments of the wind, characterized by different velocities
and hence different launching radii and mass outflow rates.

In the heartbeat state, on the other hand, we suspect that wind
is also ejected from this source, 
albeit with a density lower by about ten times and therefore undetectable from the absorption lines.
The mass outflow rates that we determined from the observation analysis 
is 2.7 and 4.2$\times 10^{17}$ g s$^{-1}$ for the two wind components observed in the stable state,
and about 2.5 and 3.9$\times 10^{16}$ g s$^{-1}$ in the heartbeat state.

From modeling the unstable accretion disk, which is partially stabilized by the wind ejected 
on the cost of the locally dissipated energy flux,
we obtained mass ouflow rates consistent with those observed in the data.
The wind outflowing as a first, faster component, is probably launched from 
the accretion disk between $\sim 380$ and 4700-5900 $R_{\rm g}$, while the second, slower component
probably originates between  $\sim 950$ and 4200-4900 Schwarzschild radii.
The mass-loss rate $\dot M_{\rm wind}$ in these two components probably is on the order
of $\lesssim 4 \times 10^{16}$ g s$^{-1}$ if the wind is weak (undetectable in our 
Chandra data) and heartbeat oscillations are still present and
on the order of  $\lesssim 4 \times 10^{17}$ g s$^{-1}$ if the wind is strong while the oscillations cease.

Now, the question remains whether the bound wind, ejected with the velocities that are below the escape velocity 
for a black hole with $6 M_{\odot}$ from the inner parts of the accretion disk, is observationally detectable. 
We suggest that
this part of the flow, also crucial for stabilizing the accretion disk oscillations, might contribute to the hard X-ray corona. 
Indeed, 
as Fig.~\ref{fig:fiamma} shows, there is no strong correlation between the flaring/nonflaring activity and the 15-50 keV hard X-ray emission. The hard X-ray emission slightly
decreases during the flaring states. 
In the current analysis, however, we cannot quantitatively 
describe this part of the flow.
The assumed time lags between the soft and hard X-rays in the case when
a moderate coronal flow develops in the flaring state might be
on the order of a second \citep{2005MNRAS.356..205J}, which means
that they are also beyond 
the scope of our current work, and it will be interesting to investigate this
 in the future.

\begin{figure}
\includegraphics[width=\columnwidth]{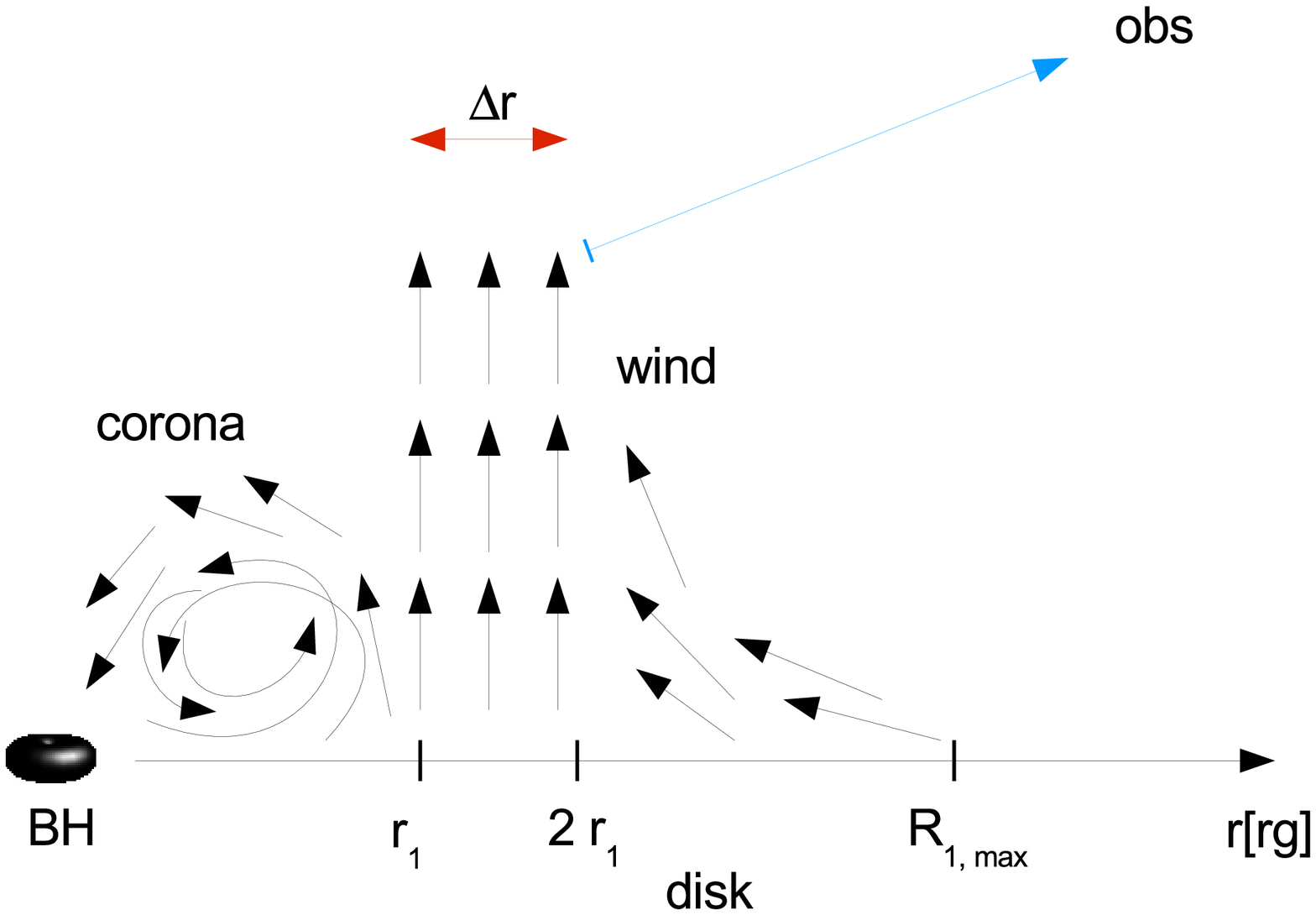}
\caption{Schematic geometry of the accretion flow in IGR J17091.
The accretion disk dominated by radiation
pressure (thin solid line) is stabilized by
a quasi-static corona in its inner parts and by the unbound wind ejected at its
outer radii (its velocity field is marked by arrows). 
The wind (we show only the first component, for clarity)
is partially collimated, and its 
radial extension in the observed wind is 
somewhat narrower than at its base. 
}
\label{fig:geometry}
\end{figure}

 In Fig. \ref{fig:geometry}, we show an overall schematic geometry of the 
accretion flow in IGR J17091. The thin accretion disk 
first develops a quasi-static corona, up to a radius $r_{1} \sim 380 r_{g}$,
where the outflow velocity exceeds the escape velocity from the BH 
gravitational potential. Above this radius, the outflow converts into the
unbound wind and is ejected from the disk up to a radius $R_{1,max} \sim 4900 r_{g}$,
with a mass-loss rate that is a function of the locally 
dissipated energy flux in the accretion disk. Moderate mass-loss 
regulates the amplitude of the disk flares that are caused by the thermal-viscous 
instability. Strong mass-loss rate suppresses these flares completely
while giving observable features in the spectra from the wind. The filling factor 
of this wind is consistent with the extension of the wind being similar
to its launching radius, that is, $\Delta r \approx r_{1}$. This implies that
the outflow is somewhat collimated already at the height where the spectral features
are produced.
In Fig. \ref{fig:geometry} only one component of the wind is shown, 
the other component, which has a different velocity and is also detected in the observations, has a similar geometry.

\subsection{Mechanisms of launching winds from accretion disks}

 Models of accretion disks postulate several different ways
of accelerating particles from the disk surface and of producing the unbound
outflows. Winds powered by magnetic fields, radiation, or thermal driving probably 
operate in active galaxies \citep{2007ASPC..373..267P}.
Wind-launching mechanisms as a result of the radiation pressure force
in their emission lines require an effective strong flux in the UV band,
which means that the distance from the black hole is determined by the temperature
of the flow. The temperature profile is sensitive to the black hole mass and
accretion rate, but this mechanism seems implausible for the
very hot accretion disks in Galactic binaries, who predominantly
radiate in the soft X-ray band.

The open magnetic field lines may be responsible for the centrifugally driven outflows
\citep{1982MNRAS.199..883B},
and the joint action of MHD and radiation pressure forces
probably plays a role in luminous AGN
\citep{2004ApJ...616..688P, 2005ApJ...631..689E}.
For the Galactic X-ray binaries, similar mechanisms of wind and jet production are
commonly suggested, with the radio luminosity being an observational
indicator of the collimated jet \citep{2003MNRAS.345.1057M}.

In the microquasar GRO J1655-40, \citet{2006Natur.441..953M} detected a 
wind component whose absorption line properties indicate that
the magnetic field is the main driving mechanism responsible for the
transfer of energy to the upper disk atmosphere layers 
as well as for the outward transfer of angular momentum in the accretion disk.
The other two other wind-driving mechanisms, the thermally driven 
wind and radiation-pressure-driven wind, are excluded. The first, based on the inverse Compton
temperature, can only occur at very large distances of $R>3\times 10^{6}$ 
$R_{Schw}$, while the observations indicated a mean radius of 200 $R_{Schw}$.
The second mechanism is excluded because the ionization parameter
for the momentum force estimated for this wind is  too large.
Eventually, therefore, we conclude that 25\% of the viscously dissipated flux, if used to 
drive the wind against the gravity force, would be enough to 
account for the observed geometry, where the height of the wind is about 0.15 $R$.

In our case, the thermally driven wind is also rather excluded.
If it was to be launched at the very large distance from the black hole,
then the regions of adequate temperature range could be outside the
disk outer edge.
We estimated the outer edge location of the accretion disk, based
on the Roche lobe size estimate given by \citet{1971ARA&A...9..183P}
 assuming the Main Sequence donor star of 
1 $M_{\odot}$, and the orbital period of 4 days \citep{2012MNRAS.422L..91W},
to be about $R_{out}\sim 4\times10^{4}$ Schwarzschild radii.

\subsection{Viability of the radiation pressure instability model}

The best-studied example of the radiation pressure instability in action 
is the microquasar GRS 1915+105, which in some spectral states exhibits 
cyclic X-ray outbursts that fit limited cycle oscillations 
of an accretion disk well. This source has been known for 20 years, and only recently another microquasar of that type was discovered. 
This is IGR J17091, the second
excellent candidate source that shows radiation-pressure-driven variability on observable timescales.

In addition, other selected Galactic X-ray binaries that were
discussed in more detail 
in \citet{2011MNRAS.414.2186J}
support that there are  $P_{rad}$ instabilities 
in their accretion disks.
Nevertheless, many sources accrete at a high accretion rate to even 
Eddington rate ratios that show no limited cycle oscillations
on timescales adequate for the radiation pressure instability.
Some stabilizing mechanisms 
need to be considered to explain the apparent stability of the high 
accretion rate sources.
One plausible mechanisms discussed in the literature that may 
affect the stability of accretion disks is the jet or wind outflow. 
Another effect that has been discussed recently
\citep{2010ApJ...718.1345K}
is the potentially 
stabilizing effect of a companion star in a binary system, 
or a companion galaxy, in case of quasars. Finally,
a physically plausible possibility is viscosity propagation throughout 
the accretion disk, as found by \citet{2012A&A...540A.114J}, which 
suppresses the oscillations in the
viscous timescale, leaving only those on the thermal timescale.

To test the latter possibility, we used our recently 
developed scheme of
propagating viscous fluctuations in the unstable disk 
\citep{2004MNRAS.348..111K,1997MNRAS.292..679L,2012A&A...540A.114J}.
The fluctuations, which propagate as in the Markov chain model,
stabilize the disk and lead to purely stochastic variability when only the
fluctuating part in the viscosity, $\alpha(r,t),$ is large.
The larger the fluctuating part, the more stochastic flickering is produced 
by the disk, and finally, no heartbeat oscillations remain. Nevertheless, 
neither of these possibilities consideres mass outflow from the disk and its correlation with   heartbeat/no-heartbeat states, 
which is clearly observed.

The luminosity outbursts for the pure radiation pressure instability 
are too large
with respect to the amplitudes observed in
IGR J17091.
The most plausible scenario is that the energetic wind is launched
from the accretion disk and partially stabilizes it,
depending on its strength and mass-loss rate. 
We determined the range of plausible wind strengths based on the timescales 
and amplitudes of the disk flares they allow.
The wind may either partially stabilize the disk oscillations to
produce regular outbursts, but of moderate amplitudes, or even
completely stabilize the disk. 

\section{Conclusions}

We studied the microquasar IGR J17091 with respect to its unique properties. It is the second known source (after GRS 1915+105) that shows the
limited-cycle oscillations characteristic for the 
thermal-viscous instability of an accretion disk. For the physical parameters of the source
that are known from observations, albeit with some uncertainties (its black hole mass
and Eddington ratio), and assuming the still-used
phenomenological $\alpha$-viscosity 
prescription of the Shakura-Sunyaev model, the hydrodynamical evolution of the disk
shows regular luminosity oscillations. Nevertheless, for the required
timescales and amplitudes of these observed oscillations, as well as to suppress
them further whenever the source enters its quasi-static mode, we need to account for
an additional physical mechanism to remove some fraction of the viscously dissipated flux and
partially (or completely) stabilize the disk.
The best process in this context is launching of the outflow from the accretion disk,
as shown not only theoretically, but also motivated by the independent spectroscopic 
observations of this source.

We  demonstrated that the wind strength  extracted from the data agrees well with our theoretical scenario. Even assuming that part of the wind is not observable because it is totally ionized, we verified that in any case
the mass outflow rate  ($3 - 4.7 \times 10^{17}$ g s$^{-1}$) of the observed wind in the no-heartbeat state is higher than the rate needed in our scenario to totally stabilize the heartbeat. 
Moreover, the lack of any lines detected in the \textit{Chandra} spectra suggests for the heartbeat
state an upper limit for the mass outflow rate 
that is compatible with the rate obtained by our model  to obtain  a flare amplitude that agrees with the observed amplitudes
($2.5 - 4 \times 10^{16}$ g s$^{-1}$).

Our final conclusions are as follows:

\begin{itemize}
\item The heartbeat oscillations of the X-ray luminosity
of IGR J17091 detected during its outburst in 2011 are attributed to the radiation pressure instability of the accretion disk.
\item The outflow launched from the disk at the cost of part of the locally dissipated
energy flux is a plausible mechanism to regulate the amplitude of these oscillations.
\item The strong outflow may stabilize the disk and completely
suppress the heartbeat.
\item The observed properties of the wind detected from IGR J17091 in the state 
without the heartbeats allowed us to constrain the the mass-loss rate at large distances 
in the disk.
\item The MHD-driven wind from the accretion disk, used in our model to stabilize the disk, is strong in the innermost parts. In the spectroscopic observations, it 
may not always be detectable because of its high ionization state or because the velocities required very close to the black hole
are below the
escape velocity; in that case, the MHD mechanism
 instead drives the formation of a quasi-static bound corona above the disk. However, we demonstrated that the observed winds agree well with the scenario described by our model.
\end{itemize}

\section*{Acknowledgments}
We thank Bozena Czerny, Daniel Proga, and Agata Rozanska
 for helpful discussions. 
This work was supported in part by the grant DEC-2012/05/E/ST9/03914 from the
Polish National Science Center.
We also acknowledge support from the COST Action MP0905,
"Black Holes in a Violent Universe".


\bibliographystyle{aa} 
\bibliography{microa} 

\end{document}